\documentclass[english]{article}
\usepackage[T1]{fontenc}
\usepackage[latin9]{inputenc}
\usepackage{color}
\usepackage{amsmath}
\usepackage{amssymb}
\usepackage{esint}

\makeatletter
\newcommand{\lyxaddress}[1]{
\par {\raggedright #1
\vspace{1.4em}
\noindent\par}
}

\makeatother

\usepackage{babel}
\begin{document}

\title{\textbf{Hawking radiation - quasi-normal modes correspondence and
effective states for nonextremal Reissner-Nordstr\"{o}m black holes}}

\author{\textbf{${\ensuremath{^{1,2,*}}}$C. Corda, $^{3,4,**}$S. H. Hendi,
$^{5,+}$R. Katebi, $^{6,++}$N. O. Schmidt}}

\maketitle

\lyxaddress{${\ensuremath{^{1}}}$Dipartimento di Fisica e Chimica, Università
Santa Rita and Institute for Theoretical Physics and Advanced Mathematics
Einstein-Galilei (IFM), 59100 Prato, Italy}

\lyxaddress{${\ensuremath{^{2}}}$International Institute for Applicable Mathematics
\& Information Sciences (IIAMIS),  Hyderabad (India) \& Udine (Italy) }

\lyxaddress{${\ensuremath{^{3}}}$Physics Department and Biruni Observatory,
College of Sciences, Shiraz University, Shiraz 71454, Iran }

\lyxaddress{${\ensuremath{^{4}}}$Research Institute for Astrophysics and Astronomy
of Maragha (RIAAM), P.O. Box 55134-441, Maragha, Iran }

\lyxaddress{${\ensuremath{^{5}}}$Department of Physics, California State University
Fullerton, 800 North State College Boulevard, Fullerton, CA 92831,
USA }

\lyxaddress{${\ensuremath{^{6}}}$Department of Mathematics, Boise State University,
1910 University Drive, Boise, ID 83725, USA }

\begin{center}
\textit{E-mail addresses:} \textcolor{blue}{$^{*}$cordac.galilei@gmail.com,
$^{**}$hendi@shirazu.ac.ir, $^{+}$rkatebi.gravity@gmail.com, $^{++}$nathanschmidt@u.boisestate.edu }
\par\end{center}
\begin{abstract}
It is known that the nonstrictly thermal character of the Hawking
radiation spectrum harmonizes Hawking radiation with black hole (BH)
quasi-normal modes (QNM). This paramount issue has been recently analyzed
in the framework of both Schwarzschild BHs (SBH) and Kerr BHs (KBH). In this assignment,
we generalize the analysis to the framework of \emph{nonextremal}
Reissner-Nordstr\"{o}m BHs (RNBH). Such a generalization is important
because in both SBHs and KBHs an absorbed (or emitted) particle
has only mass. Instead, in RNBHs the particle has charge as well as
mass. In doing so, we expose that for the RNBH, QNMs can be naturally interpreted
in terms of quantum levels for both particle emission and absorption.
Conjointly, we generalize some concepts concerning the RNBH's ``effective
states''. 
\end{abstract}

\section{Introduction}

A RNBH of mass $M$ is identical to a SBH of mass
$M$ except that a RNBH has the \emph{nonzero} charge quantity $Q$.
In this paper, we are interested in RNBHs with the \emph{nonextremal}
constraint $M>Q$ \cite{key-1}. The quantity $Q$ is the physical
mechanism for the RNBH's \emph{dual} horizons from eq. (1) in \cite{key-1}
\begin{equation}
r_{\pm}=R_{\pm_{RNBH}}(M,Q)=M\pm\sqrt{M^{2}-Q^{2}},\label{eq:RNBH-horizons}
\end{equation}
because the RNBH outer (event) horizon radius $R_{+_{RNBH}}(M,Q)$
and the \emph{RNBH inner (Cauchy) horizon radius} $R_{-_{RNBH}}(M,Q)$
are clearly functions of both $M$ \emph{and} $Q$, not just $M$
as in the well known case of the \emph{SBH horizon radius} 
\begin{equation}
r_{s}=R_{SBH}(M)=2M.
\end{equation}
Energy conservation plays a fundamental role in BH radiance \cite{key-2}
because the emission or absorption of a Hawking quanta with mass $m$
and energy-frequency $\omega$ causes a BH of mass $M$ to undergo
a transition between \emph{discrete} energy spectrum levels \cite{key-3}-\cite{key-7},
where 
\begin{equation}
E=m=\omega=\Delta M\label{eq:mass-energy-frequency-equivalence}
\end{equation}
for $G=c=k_{B}=\hbar=\frac{1}{4\pi\epsilon_{0}}=1$ (Planck units).
Given that emission and absorption are \emph{reverse} processes for
the quantized energy spectrum conservation \cite{key-3}-\cite{key-7},
we consider this pair of transitions as being \emph{equal in magnitude}
but \emph{opposite in direction} from the neutral radius perspective
of $r_{0}=(r_{+}+r_{-})/2$.

It is known that the countable character of successive emissions of
Hawking quanta which is a consequence of the nonstrictly thermal
character of the Hawking radiation spectrum (see \cite{key-3}-\cite{key-7}
and \cite{key-22}-\cite{key-26}), generates a natural correspondence
between Hawking radiation and BH QNM \cite{key-3}-\cite{key-7}.
Moreover, it has also been shown that QNMs can be naturally interpreted
in terms of quantum levels, where the emission or absorption of a
particle is interpreted as a transition between two distinct levels
on the \emph{discrete} energy spectrum \cite{key-3}-\cite{key-7}.
The thermal spectrum correction is an imperative adjustment to the
physical interpretation of BH QNMs because these results are important
to realize the underlying unitary quantum gravity theory \cite{key-3}-\cite{key-7}.
Hod's intriguing works \cite{key-8,key-9} suggested that BH QNMs
carry principle information regarding a BH's horizon area quantization.
Hod's influential conjecture was later refined and clarified by Maggiore
\cite{key-10}. Moreover, it is also believed that QNMs delve into
the micro structure of spacetime \cite{key-11}. 

To make sense of the state space for the energy spectrum states and
the underlying BH perturbation field states, an \emph{effective framework}
based on the \emph{nonstrictly thermal} behavior of Hawking's framework
began to emerge \cite{key-3}-\cite{key-7}. In the midst of this
superceding BH effective framework \cite{key-3}-\cite{key-7}, the
BH effective state concept was originally introduced for KBHs
in \cite{key-6} and subsequently applied through Hawking's
periodicity arguments \cite{key-30,key-31}, to the BH tunneling mechanism's
nonstrictly black body spectrum \cite{key-7}. The effective state
is meaningful to BH physics and thermodynamics research because one
needs additional features and knowledge to consider in future experiments
and observations.

In this paper, our objective is to apply the nonstrictly thermal
BH effective framework of \cite{key-3}-\cite{key-7} to nonextremal
RNBHs. Thus, upon recalling that a RNBH of mass $M$ is identical
to a SBH of mass $M$ except that a RNBH has the charge $Q$, we prepare
for our BH QNM investigation by reviewing relevant portions of the
SBH effective framework \cite{key-3}-\cite{key-7} for quantities
related to SBH states and transitions in Section 2. Then in Section
3, we launch our RNBH QNM exploration by introducing a RNBH effective
framework for quantities pertaining to RNBH states and transitions.
Finally, we conclude with a brief comparison between the fundamental
SBH and RNBH results of Section 4 followed by the recapitulation of
Section 5.

\section{Schwarzschild black hole framework: background and review}

\subsection{Schwarzschild black hole states and transitions}

Here, we recall some quantities that characterize the SBH.

First, consider a SBH of initial mass $M$, when the SBH emits or
absorbs a quantum of energy-frequency $\omega$ (for particle mass
$m$ and SBH mass change $\Delta M$, such that $m=\omega=\Delta M$)
to achieve a final mass of $M-\omega$ or $M+\omega$, respectively,
for the SBH mass-energy transition between states in state space.
Thus, we follow \cite{key-3}-\cite{key-5}, where the \emph{SBH initial
and final horizon area} are 
\begin{equation}
\begin{array}{rclcl}
A_{SBH}(M) & = & 16\pi M^{2} & = & 4\pi R_{SBH}^{2}(M)\\
\\
A_{SBH}(M\pm\omega) & = & 16\pi(M\pm\omega)^{2} & = & 4\pi R_{SBH}^{2}(M\pm\omega),
\end{array}\label{eq:SBH-horizon-area}
\end{equation}
respectively, for the \emph{SBH area quanta number} 
\begin{equation}
N_{SBH}(M,\omega)=\frac{A_{SBH}(M)}{|\Delta A_{SBH}(M,\omega)|},
\end{equation}
such that the \emph{SBH horizon area change} for the corresponding
mass change $\Delta M$ is 
\begin{equation}
\begin{array}{rcl}
\Delta A_{SBH}(M,\omega) & = & A_{SBH}(M\pm\omega)-A_{SBH}(M)\\
 & = & 32\pi M\omega+O(\omega^{2})\sim32\pi M\Delta M\\
 & = & 32\pi M\Delta E,
\end{array}\label{eq:SBH-horizon-area-change}
\end{equation}
because the transition's minus ($-$) and plus ($+$) signs depend
on emission and absorption, respectively. Next, in \cite{key-3}-\cite{key-5},
the \emph{Bekenstein-Hawking SBH initial and final entropy} are 
\begin{equation}
\begin{array}{rcl}
S_{SBH}(M) & = & \frac{A_{SBH}(M)}{4}\\
\\
S_{SBH}(M\pm\omega) & = & \frac{A_{SBH}(M\pm\omega)}{4},
\end{array}\label{eq:SBH-entropy}
\end{equation}
respectively, where the corresponding \emph{SBH entropy change} is
\begin{equation}
\Delta S_{SBH}(M,\omega)=\frac{\Delta A_{SBH}(M,\omega)}{4}.\label{eq:SBH-entropy-change}
\end{equation}
Subsequently, the \emph{SBH initial and final total entropy} are \cite{key-3}-\cite{key-5}
\begin{equation}
\begin{array}{rcl}
S_{SBH-total}(M) & = & S_{SBH}(M)-\ln S_{SBH}(M)+\frac{3}{2A_{SBH}(M)}\\
\\
S_{SBH-total}(M\pm\omega) & = & S_{SBH}(M\pm\omega)-\ln S_{SBH}(M\pm\omega)+\frac{3}{2A_{SBH}(M\pm\omega)},
\end{array}\label{eq:SBH-total-entropy}
\end{equation}
 respectively. Additionally, the \emph{SBH initial and final Hawking
temperature} are \cite{key-3}-\cite{key-5} 
\begin{equation}
\begin{array}{rcl}
T_{H_{SBH}}(M) & = & \frac{1}{8\pi M}\\
\\
T_{H_{SBH}}(M\pm\omega) & = & \frac{1}{8\pi(M\pm\omega)},
\end{array}\label{eq:SBH-hawking-temperature}
\end{equation}
 respectively. Therefore, the quantum transition's \emph{SBH emission
tunneling rate} is \cite{key-3}-\cite{key-5} 
\begin{equation}
\begin{array}{rcl}
\Gamma_{SBH}(M,\omega) & \sim & \exp\left[-8\pi M\omega\left(1-\frac{\omega}{2M}\right)\right]\\
 & \sim & \exp\left[-\frac{\omega}{T_{H_{SBH}}(M)}\left(1-\frac{\omega}{R_{SBH}(M)}\right)\right]\\
 & \sim & \exp\left[+\Delta S_{SBH}(M,\omega)\right].
\end{array}\label{eq:SBH-emission-tunneling-rate}
\end{equation}

\subsection{Schwarzschild black hole effective states and transitions}

Here, we recall some effective quantities that characterize the SBH.

Given that $M$ is the mass state \emph{before} and $M\pm\omega$
is the mass state \emph{after} the quantum transition, the \emph{SBH
effective mass} and \emph{SBH effective horizon} are respectively
identified in \cite{key-3}-\cite{key-5} as 
\begin{equation}
\begin{array}{rcl}
M_{E}(M,\omega) & = & \frac{M+(M\pm\omega)}{2}=M\pm\frac{\omega}{2}\\
\\
R_{E_{SBH}}(M,\omega) & = & 2M_{E}(M,\omega)
\end{array},\label{eq:SBH-effective-mass-and-horizon}
\end{equation}
which are \emph{average quantities} between the two states \emph{before}
and \emph{after} the process \cite{key-3}-\cite{key-5}. Consequently,
using eqs. ($\ref{eq:SBH-horizon-area}$) and ($\ref{eq:SBH-effective-mass-and-horizon}$)
we define the \emph{SBH effective horizon area} as 
\begin{equation}
\begin{array}{c}
A_{E_{SBH}}(M,\omega)\equiv\frac{A_{SBH}(M)+A_{SBH}(M\pm\omega)}{2}=\\
\\
\equiv16\pi M_{E}^{2}(M,\omega)=4\pi R_{E_{SBH}}^{2}(M,\omega),
\end{array}\label{eq:SBH-effective-horizon-area}
\end{equation}
which is the average of the SBH's initial and final horizon areas.
Subsequently, utilizing eq. ($\ref{eq:SBH-entropy}$), the Bekenstein-Hawking
\emph{SBH effective entropy} is defined as 
\begin{equation}
S_{E_{SBH}}(M,\omega)\equiv\frac{S_{SBH}(M)+S_{SBH}(M\pm\omega)}{2},\label{eq:SBH-effective-entropy}
\end{equation}
and consequently employs eqs. ($\ref{eq:SBH-effective-horizon-area}$)
and ($\ref{eq:SBH-effective-entropy}$) to define the \emph{SBH effective
total entropy} as 
\begin{equation}
\begin{array}{c}
S_{E_{SBH-total}}(M,\omega)\equiv\\
\\
\equiv S_{E_{SBH}}(M,\omega)-\ln S_{E_{SBH}}(M,\omega)+\frac{3}{2A_{E_{SBH}}(M,\omega)}.
\end{array}\label{eq:SBH-effective-total-entropy}
\end{equation}
Thus, employing eqs. (\emph{$\ref{eq:mass-energy-frequency-equivalence}$})
and ($\ref{eq:SBH-hawking-temperature}$), the \emph{SBH effective
temperature} is \cite{key-3}-\cite{key-5} 
\begin{equation}
\begin{array}{rcl}
T_{E_{SBH}}(M,\omega) & = & \left(\frac{T_{H_{SBH}}^{-1}(M)+T_{H_{SBH}}^{-1}(M\pm\omega)}{2}\right)^{-1}\\
 & = & \left(8\pi\left[\frac{M+M\pm\omega}{2}\right]\right)^{-1}\\
 & = & \frac{1}{4\pi(2M\pm\omega)}=\frac{1}{8\pi M_{E}(M,\omega)},
\end{array}\label{eq:SBH-effective-temperature}
\end{equation}
which is the inverse of the average value of the inverses of the initial
and final Hawking temperatures. Consequently, eq. ($\ref{eq:SBH-effective-temperature}$)
lets one rewrite eq. ($\ref{eq:SBH-emission-tunneling-rate}$) to
define the \emph{SBH effective emission tunneling rate} (in the Boltzmann-like
form) as \cite{key-3}-\cite{key-5} 
\begin{equation}
\Gamma_{E_{SBH}}(M,\omega)\sim\exp\left[-\frac{\omega}{T_{E_{SBH}}(M,\omega)}\right]=\exp\left[+\Delta S_{E_{SBH}}(M,\omega)\right],\label{eq:SBH-emission-tunneling-rate-Boltzman}
\end{equation}
such that eq. ($\ref{eq:SBH-effective-entropy}$) defines the \emph{SBH
effective entropy change} as 
\begin{equation}
\Delta S_{E_{SBH}}(M,\omega)=S_{SBH}(M\pm\omega)-S_{SBH}(M)=\frac{\Delta A_{E_{SBH}}(M,\omega)}{4}\label{eq:SBH-effective-entropy-change}
\end{equation}
because the \emph{SBH effective horizon area change} is 
\begin{equation}
\Delta A_{E_{SBH}}(M,\omega)=16\pi M_{E}(M,\omega)\omega\label{eq:SBH-effective-horizon-area-change}
\end{equation}
and the \emph{SBH effective area quanta number} is 
\begin{equation}
N_{E_{SBH}}(M,\omega)=\frac{A_{E_{SBH}}(M,\omega)}{\Delta A_{E_{SBH}}(M,\omega)}.\label{eq:SBH-effective-area-quanta-number}
\end{equation}

\subsection{Effective application of quasi-normal modes to the Schwarzschild
black hole}

Here, we recall how the SBH perturbation field QNM states can be applied
to the SBH effective framework.

The quasi-normal frequencies (QNF) are typically labeled as $\omega_{nl}$,
where $l$ is the angular momentum quantum number \cite{key-3}-\cite{key-5,key-10,key-12}.
Thus, for each $l$, such that $l\ge2$ for gravitational perturbations,
there is a countable sequence of QNMs labeled by the overtone number
$n$, which is a natural number \cite{key-3}-\cite{key-5,key-10}.

Now $|\omega_{n}|$ is the damped harmonic oscillator's proper frequency
that is defined as \cite{key-3}-\cite{key-5,key-10} 
\begin{equation}
|\omega_{n}|=(\omega_{0})_{n}=\sqrt{\omega_{n_{\mathbb{R}}}^{2}+\omega_{n_{\mathbb{I}}}^{2}}.\label{eq:SBH-QNF-generic-pythagorean}
\end{equation}
Maggiore \cite{key-10} articulated that the establishment $|\omega_{n}|=\omega_{n_{\mathbb{R}}}$
is only correct for the very long-lived and lowly excited QNMs approximation
$|\omega_{n}|\gg\omega_{n_{\mathbb{I}}}$, whereas for a lot of BH
QNMs, such as those that are highly excited, the opposite limit is
correct \cite{key-3}-\cite{key-5,key-10}. Therefore, the $\omega$
parameter in eqs. ($\ref{eq:SBH-effective-mass-and-horizon})-(\ref{eq:SBH-effective-area-quanta-number}$)
is substituted for the $|\omega_{n}|$ parameter \cite{key-3}-\cite{key-5}
because we wish to employ BH QNFs. When $n$ is large the SBH QNFs
become independent of $l$ and thereby exhibit the nonstrictly thermal
structure \cite{key-3}-\cite{key-5} 
\begin{equation}
\begin{array}{rcl}
\omega_{n} & = & \ln3\times T_{E_{SBH}}(M,|\omega_{n}|)+2\pi i(n+\frac{1}{2})\times T_{E_{SBH}}(M,|\omega_{n}|)+\mathcal{O}(n^{-\frac{1}{2}})\\
\\
 & = & \frac{\ln3}{4\pi\left[2M-|\omega_{n}|\right]}+\frac{2\pi i}{4\pi\left[2M-|\omega_{n}|\right]}(n+\frac{1}{2})+\mathcal{O}(n^{-\frac{1}{2}})\\
\\
 & = & \frac{\ln3}{8\pi M_{E}(M,|\omega_{n}|)}+\frac{2\pi(n+\frac{1}{2})}{8\pi M_{E}(M,|\omega_{n}|)}i+\mathcal{O}(n^{-\frac{1}{2}}),
\end{array}\label{eq:SBH-QNF}
\end{equation}
where 
\begin{equation}
\begin{array}{rclcl}
m_{n} & \equiv & \omega_{n_{\mathbb{R}}} & = & \frac{\ln3}{8\pi M_{E}(M,|\omega_{n}|)}\\
\\
p_{n} & \equiv & \omega_{n_{\mathbb{I}}} & = & \frac{2\pi}{8\pi M_{E}(M,|\omega_{n}|)}(n+\frac{1}{2}).
\end{array}\label{eq:SBH-QNF-components}
\end{equation}
Thus, when referring to highly excited QNMs one gets $|\omega_{n}|\approx p_{n}$
\cite{key-3}-\cite{key-5}, where the quantized levels differ from
\cite{key-10} because they are not equally spaced in exact form.
Therefore, according to \cite{key-3}-\cite{key-5}, we have 
\begin{equation}
\begin{array}{rcl}
|\omega_{n}| & = & \frac{\sqrt{(\ln3)^{2}+4\pi^{2}(n+\frac{1}{2})^{2}}}{8\pi M_{E}(M,|\omega_{n}|)}\\
 & = & T_{E_{SBH}}(M,|\omega_{n}|)\sqrt{(\ln3)^{2}+4\pi^{2}(n+\frac{1}{2})^{2}},
\end{array}\label{eq:SBH-QNM-omega-zero-n}
\end{equation}
which is solved to yield 
\begin{equation}
|\omega_{n}|=M-\sqrt{M^{2}-\frac{\sqrt{(\ln3)^{2}+4\pi^{2}(n+\frac{1}{2})^{2}}}{4\pi}}\label{eq:SBH-QNF-QNF-solution}
\end{equation}
when we obey $|\omega_{n}|<M$ because a BH cannot emit more energy
than its total mass.

\section{Reissner-Nordstr\"{o}m black hole framework: an introduction}

We note that for this framework, we consider the \emph{RNBH event
horizon} features, which are derived from the $R_{+_{RNBH}}(M,Q)$
in eq. ($\ref{eq:RNBH-horizons}$).

\subsection{Reissner-Nordstr\"{o}m black hole states and transitions}

Here, we recall some quantities that characterize the RNBH.

First, consider a RNBH of initial mass $M$ and initial charge $Q$.
Using eq. ($\ref{eq:RNBH-horizons}$), we define the \emph{RNBH initial
event horizon area} as 
\begin{equation}
A_{{+}_{RNBH}}(M,Q)=4\pi(M+\sqrt{M^{2}-Q^{2}})^{2}=4\pi R_{+_{RNBH}}^{2}(M,Q),\label{eq:RNBH-outer-horizon-area}
\end{equation}
 the \emph{Bekenstein-Hawking RNBH initial entropy} as 
\begin{equation}
S_{+_{RNBH}}(M,Q)=\frac{A_{+_{RNBH}}(M,Q)}{4},\label{eq:RNBH-outer-entropy}
\end{equation}
and the \emph{RNBH initial electrostatic potential} as 
\begin{equation}
\Phi_{+}(M,Q)=\frac{Q}{4\pi R_{+_{RNBH}}(M,Q)}=\frac{Q}{4\pi(M+\sqrt{M^{2}-Q^{2}})}.\label{eq:RNBH-outer-electrostatic-potential}
\end{equation}
Consequently, eq. (17) of \cite{key-2} identifies the \emph{RNBH
initial Hawking temperature} as 
\begin{equation}
\begin{array}{c}
T_{{+H}_{RNBH}}(M,Q)=\frac{\sqrt{M^{2}-Q^{2}}}{2\pi(M+\sqrt{M^{2}-Q^{2}})^{2}}=\\
\\
=\frac{R_{+_{RNBH}}(M,Q)-R_{-_{RNBH}}(M,Q)}{A_{{+RNBH}}(M,Q)}.
\end{array}\label{eq:RNBH-initial-hawking-temperature}
\end{equation}
Second, consider when the RNBH emits or absorbs a quantum of energy-frequency
$\omega$ with charge $q$ to achieve a final mass of $M-\omega$
or $M+\omega$ and a final charge of $Q-q$ or $Q+q$, respectively,
for the RNBH mass-energy transition between states in state space.
For this, all we need to do is replace the RNBH's mass and charge
parameters in eqs. ($\ref{eq:RNBH-outer-horizon-area}$) and ($\ref{eq:RNBH-initial-hawking-temperature}$).
Thus, eq. ($\ref{eq:RNBH-outer-horizon-area}$) establishes the \emph{RNBH
final event horizon area} as 
\begin{equation}
\begin{array}{rcl}
A_{{+}_{RNBH}}(M\pm\omega,Q\pm q) & = & 4\pi R_{+_{RNBH}}^{2}(M\pm\omega,Q\pm q)\\
\\
 & = & 4\pi\left((M\pm\omega)+\sqrt{(M\pm\omega)^{2}-(Q\pm q)^{2}}\right)^{2},
\end{array}\label{eq:final-RNBH-outer-horizon-area}
\end{equation}
eq. ($\ref{eq:RNBH-outer-entropy}$) presents the \emph{Bekenstein-Hawking
RNBH final entropy} as 
\begin{equation}
S_{+_{RNBH}}(M\pm\omega,Q\pm q)=\frac{A_{+_{RNBH}}(M\pm\omega,Q\pm q)}{4},\label{eq:final-RNBH-outer-entropy}
\end{equation}
and eq. ($\ref{eq:RNBH-outer-electrostatic-potential}$) defines the
\emph{RNBH final electrostatic potential} as 
\begin{equation}
\begin{array}{rcl}
\Phi_{+}(M\pm\omega,Q\pm q) & = & \frac{Q}{4\pi R_{+_{RNBH}}(M\pm\omega,Q\pm q)}\\
\\
 & = & \frac{Q}{4\pi\left((M\pm\omega)+\sqrt{(M\pm\omega)^{2}-(Q\pm q)^{2}}\right)}
\end{array}\label{eq:final-RNBH-outer-electrostatic-potential}
\end{equation}
for usage in eq. (29) of \cite{key-13}, where it is proposed that
the \emph{RNBH adiabatic invariant} is 
\begin{equation}
\begin{array}{c}
I_{+_{RNBH}}(M,\omega,Q,q)=\int\frac{\omega-\Phi_{+}(M\pm\omega,Q)q}{\omega}=\\
\\
=\int\frac{\Delta M-\Phi_{+}(M\pm\Delta M,Q)\Delta Q}{\Delta M}
\end{array}\label{eq:outer-RNBH-AI-integral}
\end{equation}
because $\Delta Q=q$. Hence, eq. ($\ref{eq:RNBH-initial-hawking-temperature}$)
identifies the \emph{RNBH final Hawking temperature} as 
\begin{equation}
\begin{array}{rcl}
T_{{+H}_{RNBH}}(M\pm\omega,Q\pm q) & = & \frac{R_{+_{RNBH}}(M\pm\omega,Q\pm q)-R_{-_{RNBH}}(M\pm\omega,Q\pm q)}{A_{{+RNBH}}(M\pm\omega,Q\pm q)}\\
\\
 & = & \frac{\sqrt{(M\pm\omega)^{2}-(Q\pm q)^{2}}}{2\pi\left((M\pm\omega)+\sqrt{(M\pm\omega)^{2}-(Q\pm q)^{2}}\right)^{2}}.
\end{array}\label{eq:RNBH-final-hawking-temperature}
\end{equation}
Next, upon generalizing eq. (16) in \cite{key-2} and the work \cite{key-27},
we define the \emph{RNBH tunneling rate} as 
\begin{equation}
\begin{array}{c}
\Gamma_{+_{RNBH}}(M,\omega,Q,q)\sim\\
\\
\exp\left[-4\pi\left(2\omega(M\pm\frac{\omega}{2})-(M\pm\omega)\sqrt{(M\pm\omega)^{2}-(Q\pm q)^{2}}+M\sqrt{M^{2}-Q^{2}}\right)\right]\sim\\
\\
\exp\left[\Delta S_{+_{RNBH}}(M,\omega,Q,q)\right],
\end{array}\label{eq:RNBH-tunneling-rate-initial}
\end{equation}
where we utilize eq. ($\ref{eq:final-RNBH-outer-horizon-area}$) to
define the \emph{Bekenstein-Hawking RNBH entropy change} as 
\begin{equation}
\Delta S_{+_{RNBH}}(M,\omega,Q,q)=\frac{\Delta A_{{+RNBH}}(M,\omega,Q,q)}{4},\label{eq:RNBH-outer-entropy-change}
\end{equation}
such that the \emph{RNBH event horizon area change} is 
\begin{equation}
\Delta A_{+_{RNBH}}(M,\omega,Q,q)=A_{+_{RNBH}}(M\pm\omega,Q\pm q)-A_{+_{RNBH}}(M,Q)\label{eq:RNBH-outer-horizon-area-change}
\end{equation}
so we can define the \emph{RNBH event horizon area quanta number}
as 
\begin{equation}
N_{+_{RNBH}}(M,\omega,Q,q)=\frac{A_{{+}_{RNBH}}(M,Q)}{|\Delta A_{+_{RNBH}}(M,\omega,Q,q)|}.\label{eq:RNBH-outer-area-quanta-number}
\end{equation}

\subsection{Reissner-Nordstr\"{o}m black hole effective states and transitions}

Here, we define some effective quantities that characterize the RNBH.

The \emph{RNBH effective mass} is equivalent to the SBH effective
mass component of eq. ($\ref{eq:SBH-effective-mass-and-horizon}$),
which is 
\begin{equation}
M_{E}(M,\omega)\equiv\frac{M+(M\pm\omega)}{2}.\label{eq:RNBH-effective-mass}
\end{equation}
Next, we define the \emph{RNBH effective charge} as 
\begin{equation}
Q_{E}(Q,q)\equiv\frac{Q+(Q\pm q)}{2},\label{eq:RNBH-effective-charge}
\end{equation}
which is the average of the RNBH's initial charge $Q$ and final charge
$Q\pm q$. From this, eqs. ($\ref{eq:RNBH-horizons}$), ($\ref{eq:RNBH-effective-mass}$)
and ($\ref{eq:RNBH-effective-charge}$) are used to define the corresponding
\emph{RNBH effective event horizon} and \emph{RNBH effective cauchy
horizon} as 
\begin{equation}
r_{\pm E}\equiv R_{{\pm E}_{RNBH}}(M,\omega,Q,q)\equiv M_{E}(M,\omega)\pm\sqrt{M_{E}^{2}(M,\omega)-Q_{E}^{2}(Q,q)},\label{eq:RNBH-effective-horizons}
\end{equation}
with respect to the energy conservation and pair production neutrality
of eq. ($\ref{eq:RNBH-effective-mass}$). Next, we employ eqs. ($\ref{eq:RNBH-outer-horizon-area}$),
($\ref{eq:RNBH-effective-mass}$) and ($\ref{eq:RNBH-effective-horizons}$)
to define the \emph{RNBH effective event horizon area} as 
\begin{equation}
\begin{array}{rcl}
A_{{+E}_{RNBH}}(M,\omega,Q,q) & \equiv & 4\pi R_{{+E}_{RNBH}}^{2}(M,\omega,Q,q)\\
\\
 & \equiv & 4\pi\left(M_{E}(M,\omega)+\sqrt{M_{E}^{2}(M,\omega)-Q_{E}^{2}(Q,q)}\right)^{2},
\end{array}\label{eq:RNBH-effective-outer-horizon-area}
\end{equation}
which is then used to define the \emph{RNBH effective entropy} as
\begin{equation}
S_{{+E}_{RNBH}}(M,\omega,Q,q)\equiv\frac{A_{{+E}_{RNBH}}(M,\omega,Q,q)}{4}.\label{eq:RNBH-effective-outer-entropy}
\end{equation}
Afterwards, we use eqs. ($\ref{eq:RNBH-outer-electrostatic-potential}$)
and ($\ref{eq:RNBH-effective-outer-horizon-area}$) to define the
\emph{RNBH effective electrostatic potential} as 
\begin{equation}
\begin{array}{rcl}
\Phi_{+E}(M,\omega,Q,q) & \equiv & \frac{Q_{E}(Q,q)}{4\pi R_{{+E}_{RNBH}}(M,\omega,Q,q)}\\
\\
 & \equiv & \frac{Q_{E}(Q,q)}{4\pi\left(M_{E}(M,\omega)+\sqrt{M_{E}^{2}(M,\omega)-Q_{E}^{2}(Q,q)}\right)}
\end{array}\label{eq:RNBH-effective-outer-electrostatic-potential}
\end{equation}
so we can utilize the $T_{E_{SBH}}(M,\omega)$ in eq. ($\ref{eq:SBH-effective-temperature}$)
along with eqs. ($\ref{eq:RNBH-effective-mass}$), ($\ref{eq:RNBH-effective-charge}$),
and ($\ref{eq:RNBH-effective-outer-electrostatic-potential}$) to
define the \emph{RNBH effective adiabatic invariant} as 
\begin{equation}
I_{{+E}_{RNBH}}(M,\omega,Q,q)\equiv\int\frac{dM_{E}(M,\omega)-\Phi_{+E}(M,\omega,Q,q)dQ_{E}(Q,q)}{T_{E_{SBH}}(M,\omega)}.\label{eq:effective-outer-RNBH-AI-integral}
\end{equation}
At this point, eqs. ($\ref{eq:SBH-effective-temperature}$) and ($\ref{eq:RNBH-tunneling-rate-initial}$)
let us introduce and define the \emph{RNBH effective temperature}
as 
\begin{equation}
\begin{array}{rcl}
T_{{+E}_{RNBH}}(M,\omega,Q,q) & \equiv & \frac{\sqrt{\left(M\pm\frac{\omega}{2}\right)^{2}-\left(Q\pm\frac{q}{2}\right)^{2}}}{2\pi\left[\left(M\pm\frac{\omega}{2}\right)+\sqrt{\left(M\pm\frac{\omega}{2}\right)^{2}-\left(Q\pm\frac{q}{2}\right)^{2}}\right]^{2}}\\
\\
 & \equiv & \frac{\sqrt{M_{E}^{2}(M,\omega)-Q_{E}^{2}(Q,q)}}{2\pi\left(M_{E}(M,\omega)+\sqrt{M_{E}^{2}(M,\omega)-Q_{E}^{2}(Q,q)}\right)^{2}}\\
\\
 & \equiv & \frac{R_{{+E}_{RNBH}}(M,\omega,Q,q)-R_{{-E}_{RNBH}}(M,\omega,Q,q)}{A_{{+E}_{RNBH}}(M,\omega,Q,q)},
\end{array}\label{eq:RNBH-effective-temperature}
\end{equation}
which authorizes us to exercise eqs. ($\ref{eq:RNBH-outer-entropy-change}$)
and ($\ref{eq:RNBH-effective-temperature}$) to rewrite eq. ($\ref{eq:RNBH-tunneling-rate-initial}$)
to define the RNBH effective tunneling rate as 
\begin{equation}
\begin{array}{rcl}
\Gamma_{{+E}_{RNBH}}(M,\omega,Q,q) & \sim & \exp\left[\frac{\pm\omega}{T_{{+E}_{RNBH}}(M,\omega,Q,q)}\right]\\
\\
 & \sim & \exp\left[\Delta S_{+_{RNBH}}(M,\omega,Q,q)\right],
\end{array}\label{eq:RNBH-effective-tunneling-rate}
\end{equation}
such that the \emph{RNBH effective entropy change} is defined as 
\begin{equation}
\Delta S_{{+E}_{RNBH}}(M,\omega,Q,q)\equiv\frac{\Delta A_{{+E}_{RNBH}}(M,\omega,Q,q)}{4}\label{eq:RNBH-effective-entropy-change}
\end{equation}
for the \emph{RNBH effective event horizon area change} 
\begin{equation}
\Delta A_{{+E}_{RNBH}}(M,\omega,Q,q)\equiv\frac{2\omega q+Q^{3}\pi}{(M^{2}-Q^{2})^{3/2}}\label{eq:RNBH-effective-horizon-area-change}
\end{equation}
and the \emph{RNBH effective event horizon area quanta number} 
\begin{equation}
N_{{+E}_{RNBH}}(M,\omega,Q,q)\equiv\frac{A_{{+E}_{RNBH}}(M,\omega,Q,q)}{|\Delta A_{{+E}_{RNBH}}(M,\omega,Q,q)|}.\label{eq:RNBH-effective-outer-area-quanta-number}
\end{equation}

\section{Effective application of quasi-normal modes to the Reissner-Nordstr\"{o}m
black hole}

Here, we explain how the RNBH perturbation field QNM states can be
applied to the RNBH effective framework.

Similarly to SBH QNFs, the RNBH QNFs become independent of $l$ for
large $n$ \cite{key-14}. Thus, for large $n$, we have two families
of the QNM: 
\begin{equation}
\omega_{n}=\ln3\times T_{{+H}_{SBH}}(M,Q)-2\pi(n+\frac{1}{2})i\times T_{{+H}_{SBH}}(M,Q)+\frac{qQ}{R_{+_{SBH}}(M,Q)}\label{eq:RNBH-QNF-generic-1}
\end{equation}
and 
\begin{equation}
\begin{array}{rcl}
\omega_{n} & = & \ln2\times T_{{+H}_{RNBH}}(M,Q)-2\pi(n+\frac{1}{2})i\times T_{{+H}_{RNBH}}(M,Q)+\frac{qQ}{R_{+_{RNBH}}(M,Q)}\\
\\
 & = & \frac{\ln2\sqrt{M^{2}-Q^{2}}}{2\pi\left(M+\sqrt{M^{2}-Q^{2}}\right)^{2}}-\frac{(n+\frac{1}{2})\sqrt{M^{2}-Q^{2}}}{\left(M+\sqrt{M^{2}-Q^{2}}\right)^{2}}i+\frac{qQ}{R_{+_{RNBH}}(M,Q)}.
\end{array}\label{eq:RNBH-QNF-generic-2}
\end{equation}
Now the approximation of eqs. ($\ref{eq:RNBH-QNF-generic-1}$) and
($\ref{eq:RNBH-QNF-generic-2}$) are only relevant under the assumption
that the BH radiation spectrum is strictly thermal \cite{key-3}-\cite{key-5}
because they both use the Hawking temperature $T_{{+H}_{RNBH}}$ in
eq. ($\ref{eq:RNBH-initial-hawking-temperature}$). Hence, to operate
in compliance of \cite{key-3}-\cite{key-5} and thereby account for
the thermal spectrum deviation of eq. ($\ref{eq:RNBH-tunneling-rate-initial}$),
we opt to select the eq. ($\ref{eq:RNBH-QNF-generic-2}$) case and
upgrade it by effectively replacing its $T_{H_{RNBH}}$ in eq. ($\ref{eq:RNBH-initial-hawking-temperature}$)
with the $T_{{+E}_{RNBH}}$ in eq. ($\ref{eq:RNBH-effective-temperature}$).
Therefore, the corrected expression for the RNBH QNFs of eq. ($\ref{eq:RNBH-QNF-generic-2}$)
that encodes the nonstrictly thermal behavior of the radiation spectrum
is defined as 
\begin{equation}
\begin{array}{rcl}
\omega_{n} & \equiv & \ln2\times T_{{+E}_{RNBH}}(M,|\omega_{n}|,Q,q)-2\pi(n+\frac{1}{2})i\times T_{{+E}_{RNBH}}(M,|\omega_{n}|,Q,q)\\
 &  & +\frac{qQ_{E}(Q,q)}{R_{{+E}_{RNBH}}(M,|\omega_{n}|,Q,q)}\\
\\
 & \equiv & \frac{\ln2\sqrt{M_{E}^{2}(M,|\omega_{n}|)-Q_{E}^{2}(Q,q)}}{2\pi\left(M_{E}(M,|\omega_{n}|)+\sqrt{M_{E}^{2}(M,|\omega_{n}|)-Q_{E}^{2}(Q,q)}\right)^{2}}-\frac{(n+\frac{1}{2})\sqrt{M_{E}^{2}(M,|\omega_{n}|)-Q_{E}^{2}(Q,q)}}{\left(M_{E}(M,|\omega_{n}|)+\sqrt{M_{E}^{2}(M,|\omega_{n}|)-Q_{E}^{2}(Q,q)}\right)^{2}}i\\
 &  & +\frac{qQ_{E}(Q,q)}{R_{{+E}_{RNBH}}(M,|\omega_{n}|,Q,q)}.
\end{array}\label{eq:RNBH-QNF-generic-effective}
\end{equation}
From eqs. ($\ref{eq:RNBH-effective-mass}$), ($\ref{eq:RNBH-effective-horizons}$),
and ($\ref{eq:RNBH-effective-temperature}$) we define the effective
quantities associated to the QNMs as 
\begin{equation}
M_{E}(M,|\omega_{n}|)\equiv\frac{M+(M-|\omega_{n}|)}{2},\label{eq:RNBH-effective-mass-QNM}
\end{equation}
 
\begin{equation}
\begin{array}{c}
r_{\pm E}\equiv R_{{\pm E}_{RNBH}}(M,|\omega_{n}|,Q,q)\\
\\
\equiv M_{E}(M,|\omega_{n}|)\pm\sqrt{M_{E}^{2}(M,|\omega_{n}|)-Q_{E}^{2}(Q,q)},
\end{array}\label{eq:RNBH-effective-outer-horizon-QNM}
\end{equation}
and 
\begin{equation}
\begin{array}{rcl}
T_{{+E}_{RNBH}}(M,|\omega_{n}|,Q,q) & \equiv & \frac{\sqrt{\left(M-\frac{|\omega_{n}|}{2}\right)^{2}-\left(Q-\frac{q}{2}\right)^{2}}}{2\pi\left[\left(M-\frac{|\omega_{n}|}{2}\right)+\sqrt{\left(M-\frac{|\omega_{n}|}{2}\right)^{2}-\left(Q-\frac{q}{2}\right)^{2}}\right]^{2}}\\
\\
 & = & \frac{\sqrt{M_{E}^{2}(M,|\omega_{n}|)-Q_{E}^{2}(Q,q)}}{2\pi\left(M_{E}(M,|\omega_{n}|)+\sqrt{M_{E}^{2}(M,|\omega_{n}|)-Q_{E}^{2}(Q,q)}\right)^{2}}\\
\\
 & = & \frac{R_{{+E}_{RNBH}}(M,|\omega_{n}|,Q,q)-R_{{-E}_{RNBH}}(M,|\omega_{n}|,Q,q)}{A_{{+E}_{RNBH}}(M,|\omega_{n}|,Q,q)},
\end{array}\label{eq:RNBH-effective-temperature-QNM}
\end{equation}
respectively, for the quantum overtone number $n$ in eq. ($\ref{eq:RNBH-QNF-generic-effective}$).
Hence, eq. ($\ref{eq:RNBH-QNF-generic-effective}$) lets us rewrite
the SBH case of eq. ($\ref{eq:SBH-QNF-components}$) to present the
RNBH case 
\begin{equation}
\begin{array}{rcl}
m_{n} & \equiv & \ln2\times T_{{+E}_{RNBH}}(M,|\omega_{n}|,Q,q)+\frac{eQ_{E}(Q,q)}{R_{{+E}_{RNBH}}(M,|\omega_{n}|,Q,q)}\\
 & = & \frac{\ln2\sqrt{M_{E}^{2}(M,|\omega_{n}|)-Q_{E}^{2}(Q,q)}}{2\pi\left(M_{E}(M,|\omega_{n}|)+\sqrt{M_{E}^{2}(M,|\omega_{n}|)-Q_{E}^{2}(Q,q)}\right)^{2}}+\frac{qQ_{E}(Q,q)}{R_{{+E}_{RNBH}}(M,|\omega_{n}|,Q,q)}\\
\\
p_{n} & \equiv & -2\pi(n+\frac{1}{2})\times T_{{+E}_{RNBH}}(M,|\omega_{n}|,Q,q)=-\frac{(n+\frac{1}{2})\sqrt{M_{E}^{2}(M,|\omega_{n}|)-Q_{E}^{2}(Q,q)}}{\left(M_{E}(M,|\omega_{n}|)+\sqrt{M_{E}^{2}(M,|\omega_{n}|)-Q_{E}^{2}(Q,q)}\right)^{2}}.
\end{array}\label{eq:RNBH-QNF-components}
\end{equation}
Thus, we recall that if $|\omega_{n}|\approx p_{n}$, then we are
referring to highly excited QNMs \cite{key-3}-\cite{key-5} . Therefore,
the SBH case of eq. ($\ref{eq:SBH-QNM-omega-zero-n}$) becomes the
RNBH case 
\begin{equation}
\begin{array}{rcl}
|\omega_{n}| & \equiv & \frac{\sqrt{M_{E}^{2}(M,|\omega_{n}|)-Q_{E}^{2}(Q,q)}\sqrt{(\ln2)^{2}-4\pi^{2}(n+\frac{1}{2})^{2}}}{2\pi\left(M_{E}(M,|\omega_{n}|)+\sqrt{M_{E}^{2}(M,|\omega_{n}|)-Q_{E}^{2}(Q,q)}\right)^{2}}+\frac{qQ_{E}(Q,q)}{R_{{+E}_{RNBH}}(M,|\omega_{n}|,Q,q)}\\
\\
 & = & T_{{+E}_{RNBH}}(M,|\omega_{n}|,Q,q)\sqrt{(\ln2)^{2}-4\pi^{2}(n+\frac{1}{2})^{2}}+\frac{qQ_{E}(Q,q)}{R_{{+E}_{RNBH}}(M,|\omega_{n}|,Q,q)}.
\end{array}\label{eq:RNBH-QNM-omega-zero-n}
\end{equation}
Hence, upon considering eqs. ($\ref{eq:RNBH-effective-charge}$) and
($\ref{eq:RNBH-effective-mass-QNM}$), one can rewrite eq. ($\ref{eq:RNBH-QNM-omega-zero-n}$)
as 
\begin{equation}
\begin{array}{rcl}
|\omega_{n}| & \equiv & \frac{\sqrt{\left(M-\dfrac{|\omega_{n}|}{2}\right)^{2}-\left(Q-\dfrac{q}{2}\right)^{2}}\sqrt{(\ln2)^{2}-4\pi^{2}(n+\frac{1}{2})^{2}}}{2\pi\left[\left(M-\dfrac{|\omega_{n}|}{2}\right)+\sqrt{\left(M-\dfrac{|\omega_{n}|}{2}\right)^{2}-\left(Q-\dfrac{q}{2}\right)^{2}}\right]^{2}}\\
 &  & +\frac{q\left(Q-\dfrac{q}{2}\right)}{\left(M-\dfrac{|\omega_{n}|}{2}\right)+\sqrt{\left(M-\dfrac{|\omega_{n}|}{2}\right)^{2}-\left(Q-\dfrac{q}{2}\right)^{2}}},
\end{array}\label{eq:RNBH-QNM-omega-zero-n-2}
\end{equation}
where the solution of eq. ($\ref{eq:RNBH-QNM-omega-zero-n-2}$) in
terms of $|\omega_{n}|$ will be the answer of $|\omega_{n}|$. Therefore,
given a quantum transition between the levels $n$ and $n-1$, we
define $|\triangle\omega_{n,n-1}|\equiv|\omega_{n}-\omega_{n-1}|$
where eqs. (41)-(45) are rewritten as 
\begin{equation}
\begin{array}{c}
r_{\pm E}\equiv R_{{\pm E}_{RNBH}}(M,|\triangle\omega_{n,n-1}|,Q,q)\equiv\\
\\
\equiv M_{E}(M,|\triangle\omega_{n,n-1}|)\pm\sqrt{M_{E}^{2}(M,|\triangle\omega_{n,n-1}|)-Q_{E}^{2}(Q,q)},
\end{array}\label{eq:RNBH-effective-inner-horizon-QNM}
\end{equation}
 
\begin{equation}
\begin{array}{c}
A_{{+E}_{RNBH}}(M,|\triangle\omega_{n,n-1}|,Q,q)\equiv4\pi R_{{+E}_{RNBH}}^{2}(M,|\triangle\omega_{n,n-1}|,Q,q)\\
\\
\equiv4\pi\left(M_{E}(M,|\triangle\omega_{n,n-1}|)+\sqrt{M_{E}^{2}(M,|\triangle\omega_{n,n-1}|)-Q_{E}^{2}(Q,q)}\right)^{2},
\end{array}\label{eq:RNBH-effective-outer-horizon-area-QNM}
\end{equation}
 
\begin{equation}
S_{{+E}_{RNBH}}(M,|\triangle\omega_{n,n-1}|,Q,q)\equiv\frac{A_{{+E}_{RNBH}}(M,|\triangle\omega_{n,n-1}|,Q,q)}{4},\label{eq:RNBH-effective-outer-entropy-QNM}
\end{equation}
 
\begin{equation}
\begin{array}{rcl}
\Phi_{+E}(M,|\triangle\omega_{n,n-1}|,Q,q) & \equiv & \frac{Q_{E}(Q,q)}{4\pi R_{{+E}_{RNBH}}(M,|\triangle\omega_{n,n-1}|,Q,q)}\\
\\
 & \equiv & \frac{Q_{E}(Q,q)}{4\pi\left(M_{E}(M,|\triangle\omega_{n,n-1}|)+\sqrt{M_{E}^{2}(M,|\triangle\omega_{n,n-1}|)-Q_{E}^{2}(Q,q)}\right)},
\end{array}\label{eq:RNBH-effective-outer-electrostatic-potential-QNM}
\end{equation}
 
\begin{equation}
\begin{array}{c}
I_{{+E}_{RNBH}}(M,|\triangle\omega_{n,n-1}|,Q,q)=\\
\\
=\int\frac{dM_{E}(M,|\triangle\omega_{n,n-1}|)-\Phi_{+E}(M,|\triangle\omega_{n,n-1}|,Q,q)dQ_{E}(Q,q)}{T_{E_{SBH}}(M,|\triangle\omega_{n,n-1}|)},
\end{array}\label{eq:effective-outer-RNBH-AI-integral-QNM}
\end{equation}
and eqs. (47)-(50) become 
\begin{equation}
\begin{array}{rcl}
\Gamma_{{+E}_{RNBH}}(M,|\triangle\omega_{n,n-1}|,Q,q) & \sim & \exp\left[\frac{\pm|\triangle\omega_{n,n-1}|}{T_{{+E}_{RNBH}}(M,|\triangle\omega_{n,n-1}|,Q,q)}\right]\\
\\
 & \sim & \exp\left[\Delta S_{+_{RNBH}}(M,|\triangle\omega_{n,n-1}|,Q,q)\right],
\end{array}\label{eq:RNBH-effective-tunneling-rate-QNM}
\end{equation}
\begin{equation}
\Delta S_{{+E}_{RNBH}}(M,|\triangle\omega_{n,n-1}|,Q,q)\equiv\frac{\Delta A_{{+E}_{RNBH}}(M,|\triangle\omega_{n,n-1}|,Q,q)}{4},\label{eq:RNBH-effective-entropy-change-QNM}
\end{equation}
 
\begin{equation}
\Delta A_{{+E}_{RNBH}}(M,|\triangle\omega_{n,n-1}|,Q,q)\equiv\frac{2|\triangle\omega_{n,n-1}|q+\pi Q^{3}}{(M^{2}-Q^{2})^{3/2}},\label{eq:RNBH-effective-horizon-area-change-QNM}
\end{equation}
 
\begin{equation}
N_{{+E}_{RNBH}}(M,|\triangle\omega_{n,n-1}|,Q,q)\equiv\frac{A_{{+E}_{RNBH}}(M,|\triangle\omega_{n,n-1}|,Q,q)}{|\Delta A_{{+E}_{RNBH}}(M,|\triangle\omega_{n,n-1}|,Q,q)|},\label{eq:RNBH-effective-outer-area-quanta-number-QNM}
\end{equation}
 respectively.

\section{A brief comparison}

Here, we will show that the SBH results of Section 2 are in fundamental
agreement with the RNBH results of Section 3 for small $Q$, where
we recall that the RNBH of mass $M$ is identical to a SBH of mass
$M$ except that a RNBH has the nonzero charge quantity $Q$.

First, for small $Q$, the SBH's $T_{E_{SBH}}(M,\omega)$ of eq. ($\ref{eq:SBH-effective-temperature}$)
is related to the RNBH's $T_{{+E}_{RNBH}}(M,\omega,Q,q)$ of eq. ($\ref{eq:RNBH-effective-temperature}$)
as 
\begin{equation}
T_{{+E}_{RNBH}}(M,\omega,Q,q)\equiv T_{E_{SBH}}(M,\omega)-\frac{3q^{2}Q^{2}}{8(2m\pm\omega)^{5}\pi}+\mathcal{O}(Q^{4},q^{4}).\label{eq:SBH-RNBH-consistent-effective-temperature}
\end{equation}
Second, for small $Q$, the SBH's $A_{E_{SBH}}(M,\omega)$ of eq.
($\ref{eq:SBH-effective-horizon-area}$) complies with the RNBH's
$A_{{+E}_{RNBH}}(M,\omega,Q,q)$ of eq. ($\ref{eq:RNBH-effective-outer-horizon-area}$)
as 
\begin{equation}
A_{{+E}_{RNBH}}(M,\omega,Q,q)\equiv A_{E_{SBH}}(M,\omega)-8\pi Q^{2}+\mathcal{O}(Q^{4}).
\label{eq:SBH-RNBH-consistent-effective-entropy}
\end{equation}
Third, for small $Q$, the SBH's $S_{E_{SBH}}(M,\omega)$ of eq. ($\ref{eq:SBH-effective-entropy}$)
corresponds with the RNBH's $S_{{+E}_{RNBH}}(M,\omega,Q,q)$ of eq.
($\ref{eq:RNBH-effective-outer-entropy}$) as 
\begin{equation}
S_{{+E}_{RNBH}}(M,\omega,Q,q)\equiv S_{E_{SBH}}(M,\omega)-2\pi Q^{2}+\mathcal{O}(Q^{4}).\label{eq:SBH-RNBH-consistent-effective-area}
\end{equation}
Fourth, for small $Q$, the SBH's QNF $|\omega_{n}|$ of eq. ($\ref{eq:SBH-QNM-omega-zero-n}$)
is consistent with the RNBH's QNF $|\omega_{n}|$ of eqs. ($\ref{eq:RNBH-QNM-omega-zero-n}$)
and ($\ref{eq:RNBH-QNM-omega-zero-n-2}$) as 
\begin{equation}
\begin{array}{rl}
 & |\omega_{n}|\equiv\frac{\sqrt{\ln2^{2}-4\pi^{2}(n+\frac{1}{2})^{2}}}{4(2M-|\omega_{n}|)\pi}+\frac{qQ}{2M-|\omega_{n}|}=\\
\\
 & \frac{3(16\pi M^{2}-16\pi|\omega_{n}|M+4\pi|\omega_{n}|^{2}+\sqrt{-(\ln2+\pi+2\pi n)(-\ln2+\pi+2\pi n)})}{8(2M-|\omega_{n}|)^{5}\pi}Q^{2}q^{2}\\
\\
 & \frac{q^{2}}{4M-|\omega_{n}|}+\mathcal{O}(Q^{4},q^{4}),
\end{array}=\label{eq:SBH-RNBH-consistent-omega-n}
\end{equation}
which can be applied to eqs. ($\ref{eq:SBH-RNBH-consistent-effective-temperature}$)-($\ref{eq:SBH-RNBH-consistent-effective-entropy}$)
by replacing the $\omega$ parameter with the pertinent $|\omega_{n}|$.
Hence, eqs. ($\ref{eq:SBH-RNBH-consistent-effective-temperature}$)-($\ref{eq:SBH-RNBH-consistent-omega-n}$)
indicate that in general, the SBH results of Section 2 are fundamentally
consistent with the RNBH results of Section 3 for small $Q$. Moreover,
in eq. ($\ref{eq:SBH-RNBH-consistent-omega-n}$) for large $n$, the
result is consistent with the SBH because $\ln2$ is negligible, but
for small $n$ there is an argument between scientists regarding $\ln2$
and $\ln3$ because these refer to the two distinct QNM families of
eqs. (51) and (52).

Here, we provide the physical answer of eq. ($\ref{eq:SBH-RNBH-consistent-omega-n}$)
for the case of emission by using the fact that $Q$ is small, so
the term which includes $Q^{2}$ is also very small and therefore
negligible: 
\begin{equation}
(\omega_{0})_{n}\equiv|\omega_{n}|\approx M-\sqrt{M^{2}+\frac{q^{2}}{2}-Qq-\frac{1}{4\pi}\sqrt{\ln2^{2}-4\pi^{2}(n+\frac{1}{2})^{2}}}.\label{eq:Omega-n answer}
\end{equation}
Thus, by setting $(\omega_{0})_{n}\equiv|\omega_{n}|$ we obtain 
\begin{equation}
\begin{array}{rcl}
\triangle M_{n} & \equiv-\triangle\omega_{n,n-1}= & (\omega_{0})_{n-1}-(\omega_{0})_{n}\\
 & \equiv & \sqrt{M^{2}+\frac{q^{2}}{2}-Qq-\frac{1}{4\pi}\sqrt{(\ln2)^{2}+4\pi^{2}(n+\frac{1}{2})^{2}}}\\
 &  & -\sqrt{M^{2}+\frac{q^{2}}{2}-Qq-\frac{1}{4\pi}\sqrt{(\ln2)^{2}+4\pi^{2}(n-\frac{1}{2})^{2}}}
\end{array}\label{eq: variazione 2}
\end{equation}
for an emission involving quantum levels $n$ and $n-1$, which becomes
\begin{equation}
\begin{array}{c}
\triangle M_{n}\approx\sqrt{M^{2}+\frac{q^{2}}{2}-Qq-\frac{1}{2}(n+\frac{1}{2})}-\sqrt{M^{2}+\frac{q^{2}}{2}-Qq-\frac{1}{2}(n-\frac{1}{2})}\end{array}\label{eq: variazione 3}
\end{equation}
 for large $n$.

\section{Conclusion remarks}

We began our paper by summarizing some basic similarities and differences
between SBHs and RNBHs in terms of charge and horizon radii. Moreover,
we briefly explored the Parikh-Wilczek statement that explains how
energy conservation and pair production \cite{key-2,key-21} are fundamentally
related to such BHs. For a BH's discrete energy spectrum, the emission
or absorption of a particle yields a transition between two distinct
levels, where particle emission and absorption are reverse processes
\cite{key-3}-\cite{key-7}. For this, we touched on the important
issue that the nonstrictly thermal character of Hawking's radiation
spectrum generates a natural correspondence between Hawking's radiation
and BH QNMs, because these structures exemplify features of the BH's
energy spectrum \cite{key-3}-\cite{key-5}, which has been recently
generalized to the emerging concept of a BH's effective state \cite{key-6,key-7}.

Next, we prepared for our nonextremal RNBH QNM investigation by first
reviewing relevant portions of the SBH effective framework \cite{key-3}-\cite{key-5}
in Section 2. Here, we listed the noneffective and effective quantities
for SBH states and transitions, with direct application to the QNM
characterization and framework of \cite{key-3}-\cite{key-5}. Subsequently,
in Section 3, we identified some existing noneffective quantities
and introduced new effective quantities for RNBH states and transitions
so we could apply the BH framework of \cite{key-3}-\cite{key-5}
to implement a RNBH framework. These results are crucial because the
effective quantities in \cite{key-3}-\cite{key-5} have been achieved
for the stable four dimensional RNBH solution in Einstein's general
relativity---now effective frameworks exist for the SBH, KBH, and
(nonextremal) RNBH solutions. 

Ultimately, the RNBH effective quantities permitted us to utilize
both the KBH's effective state concept \cite{key-6,key-7} and the
BH QNMs \cite{key-3}-\cite{key-5} to construct a foundation for
the RNBH's effective state in this developing BH effective framework.
The RNBH effective state concept is meaningful because, as scientists
who wish to demystify the BH paradigm, we need additional features
and knowledge to consider in future experiments and observations.

Finally, we stress that the nonstrictly thermal behavior of the Hawking
radiation spectrum has been recently used to construct two very intriguing
proposals to solve the BH information loss paradox. The first one
received the First Award in the 2013 Gravity Research Foundation Essay
Competition \cite{key-26}. The latter won the Community Rating at
the 2013 FQXi Essay Contest - It from Bit or Bit from It \cite{key-28}.
We are working to extend this second approach to the RNBH framework
\cite{key-29}.

\section*{Acknowledgements}

S. H. Hendi wishes to thank the Shiraz University Research Council.
The work of S. H. Hendi has been supported financially by the Research
Institute for Astronomy \& Astrophysics of Maragha (RIAAM), Iran.

\section*{Conflict of Interests}

The authors declare that there is no conflict of interests regarding
the publication of this article.


\begin{thebibliography}{10}
\bibitem{key-1}B. Wang, R. K. Su, P. K. N. Yu and E. C. M. Young,
Phys. Rev. D 57, 5284 (1998).

\bibitem{key-2}M. K. Parikh and F. Wilczek, Phys. Rev. Lett. 85,
5042 (2000). 

\bibitem{key-3}C. Corda, JHEP 1108, 101 (2011).

\bibitem{key-4}C. Corda, Eur. Phys. J. C 73, 2665 (2013).

\bibitem{key-5}C. Corda, Int. Journ. Mod. Phys. D 21, 1242023 (2012,
Honorable Mention at Gravity Research Foundation).

\bibitem{key-6}C. Corda, S. H. Hendi, R. Katebi, and N. O. Schmidt,
JHEP 6, 8 (2013).

\bibitem{key-7}C. Corda, Ann. Phys. 337, 49 (2013).

\bibitem{key-8}S. Hod, Gen. Rel. Grav. 31, 1639 (1999, Fifth Award
at Gravity Research Foundation). 

\bibitem{key-9}S. Hod, Phys. Rev. Lett. 81 4293 (1998). 

\bibitem{key-10}M. Maggiore, Phys. Rev. Lett. 100, 141301 (2008).

\bibitem{key-11}M. Casals, A. C. Ottewill, Phys. Rev. D 86, 024021
(2012).

\bibitem{key-12}T. Padmanabhan, CQG 21, L1 (2004).

\bibitem{key-13}A. Lopez-Ortega, CQG 28, 035009 (2011).

\bibitem{key-14}S. Hod, CQG 23, L23 (2006).

\bibitem{key-15}H. P. Nollert, Phys. Rev. D 47, 5253 (1993).

\bibitem{key-16}N. Andersson, CQG 10, L61 (1993).

\bibitem{key-17}L. Motl, Adv. Theor. Math. Phys. 6, 1135 (2003).

\bibitem{key-18}L. Motl and A. Neitzke, Adv. Theor. Math. Phys. 7,
307 (2003).

\bibitem{key-19}S. Hod, Phys. Lett. B 710, 349 (2012).

\bibitem{key-20}R. A. Konoplya and A. Zhidenko, Rev. Mod. Phys. 83,
793 (2011).

\bibitem{key-21}M. K. Parikh, Gen. Rel. Grav. 36, 2419 (2004, First
Award at Gravity Research Foundation).

\bibitem[22]{key-22}B. Zhang, Q.-Y. Cai, L. You,, and M. S. Zhan,
Phys. Lett. B 675, 98 (2009). 

\bibitem[23]{key-23}B. Zhang, Q.-Y. Cai, M. S. Zhan, and L. You,
Ann. Phys. 326, 350 (2011). 

\bibitem[24]{key-24}B. Zhang, Q.-Y. Cai, M. S. Zhan, and L. You,
EPL 94, 20002 (2011). 

\bibitem[25]{key-25}B. Zhang, Q.-Y. Cai, M. S. Zhan, and L. You,
arXiv:1210.2048.

\bibitem[26]{key-26}B. Zhang, Q.-Y. Cai, M. S. Zhan, and L. You,
to appear in a Special Issue of Int. Journ. Mod. Phys. D (2013, First
Award in the Gravity Research Foundation Essay Competition).

\bibitem[27]{key-27}J. Zhang and Z. Zhao, JHEP 10, 55 (2005).

\bibitem[28]{key-28}C. Corda, http://fqxi.org/community/forum/topic/1856.

\bibitem[29]{key-29}C. Corda, S. H. Hendi, R. Katebi, and N. O. Schmidt,
in preparation.

\bibitem[30]{key-30}R. Banerjee and B. R. Majhi, Phys. Lett. B 674,
218 (2009).

\bibitem[31]{key-31}S. W. Hawking, \textquotedblleft{}The Path Integral
Approach to Quantum Gravity\textquotedblright{}, in General Relativity:
An Einstein Centenary Survey, eds. S.W.Hawking and W.Israel, (Cambridge
University Press, 1979). \end{thebibliography}
\end{document}